\documentclass[twocolumn]{aa}
\usepackage{epsfig}
\usepackage{graphics,color}
\usepackage{natbib}

\begin{document}
   \title{In-situ acceleration of subrelativistic electrons in the Coma
 halo and the halo's influence on the Sunyaev-Zeldovich effect}


   \author{V. A. Dogiel\inst{1,2}, S. Colafrancesco\inst{3},
   C.M. Ko\inst{1}, P.H. Kuo\inst{1}, C.Y. Hwang\inst{1}, W.H. Ip\inst{1}, M. Birkinshaw\inst{4},
   D.A. Prokhorov\inst{5}
          }

 \offprints{S. Colafrancesco}

\institute{Institute of Astronomy, National Central University, Chung-Li 32054,
Taiwan
  \and
I.E.Tamm Theoretical Physics Division, P.N.Lebedev Physical Institute, 117924
Moscow, Russia; Email: dogiel@lpi.ru \and
  INAF - Osservatorio Astronomico di Roma, via Frascati 33, 00040 Monteporzio,
Italy
  \and
 Department of Physics, University of Bristol, Tyndall Avenue, Bristol BS8 1TL, UK
  \and
  Moscow Institute of Physics and Technology, Institutskii lane, 141700 Moscow Region,
  Dolgoprudnii, Russia
 }

   \date{Received August 23 2005 ; accepted September 19 2006}

\authorrunning {Dogiel et al.}
\titlerunning {In-situ acceleration of subrelativistic electrons}



\abstract
   {}
   {The stochastic acceleration of subrelativistic electrons from a background
plasma is studied in order to find a possible explanation of the hard X-ray
emission detected from the Coma cluster. }
  {   We calculate the necessary energy supply as a function of the plasma temperature
   and of the electron energy, and we show that, for the same value of the hard X-ray flux,
   the energy supply changes gradually from its high value for the case when emitting particle
   are non-thermal to lower values when the electrons are thermal.
   The kinetic equations we use include terms describing particle
 thermalization as well as momentum diffusion due to the Fermi II acceleration.}
   {We show that the temporal evolution of the particle distribution function has, at its
final stationary stage, a rather specific form. This distribution function cannot be
described by simple exponential or power-law expressions. A broad transfer region is
formed by Coulomb collisions at energies between the Maxwellian and power-law parts of
the distribution functions. In this region the radiative lifetime of a single
quasi-thermal electron differs greatly from the lifetime of the distribution function as
a whole. For a plasma temperature of 8~keV, the particles emitting bremsstrahlung at $20
- 80$~keV lie in this quasi-thermal regime. We show that the energy supply required by
quasi-thermal electrons to produce the observed hard X-ray flux from Coma is one or two
orders of magnitude smaller than the value derived from the assumption of a nonthermal
origin of the emitting particles. This result may solve the problem of rapid cluster
overheating by nonthermal electrons raised by Petrosian (2001): while Petrosian's
estimates are correct for nonthermal particles, they are inapplicable in the
quasi-thermal range. We finally analyze the change in Coma's Sunyaev-Zeldovich effect
caused by the implied distortions of the Maxwellian spectrum of electrons,  and we show
that evidence for the acceleration of subrelativistic electrons can, in principle, be
derived from detailed spectral measurements.}
   {}


   \maketitle

\section{Introduction}\label{sec:intro}

An excess of hard X-ray (hereafter HXR) emission above the thermal spectrum has
been found in the energy range $20 - 80$~keV from the Coma cluster of galaxies
\citep{fus99, rep02, fus04}. The validity of this excess is, however, still
unclear because \citet{ross04} re-analyzed the Coma data and found no evidence
for such hard X-ray excess. Further re-analysis of the same data \citep{fus04}
seem to confirm the presence of the HXR excess at the level observed by
\citet{fus99} and \citet{rep02}. The recent INTEGRAL observation of Coma
(Renaud et al. 2006) analyzed the morphology of the cluster in the range 18-30
keV and found that the hard X-ray emission comes from an extended  source with
a radius $\sim 30$ arcmin. The spatial distribution is similar to the thermal
one as obtained with XMM (in the range  0.3-2 keV). The INTEGRAL data indicate
that the upper limit in 30-50 keV range is a factor $\approx 1.5$ above the
mean RXTE spectrum and the non-thermal mechanisms are expected to contribute
$\sim 50 \%$ of the total flux in this region.

The hard X-ray excess has been interpreted as being due either to inverse Compton (IC)
scattering of relic CMB photons by relativistic electrons \citep[see,
e.g.,][]{sar98,fus99,bru01} or to bremsstrahlung of nonthermal subrelativistic electrons
\citep{elb99,ks00}. Yet further interpretations have been proposed
--- X-ray emission by secondary electrons \citep{bla99}, or bremsstrahlung
emission by subrelativistic protons \citep{dog01}.

Each of these models has serious problems.  In the framework of the IC model
the magnetic field strength can be estimated from the ratio of X-ray to radio
fluxes because both radiations are produced by the same relativistic electrons.
A weak (uniform) value of the magnetic field of order 0.1-0.2 $\mu$G is derived
from the IC model. On the other hand, estimates of the magnetic field strength
in the intracluster medium determined from Faraday rotation yield much higher
value of the order of $\approx 5 - 10 \ \rm \mu G$ in the cluster center
\citep[see, e.g.,][]{cla01,gov}. However, modifications of the IC model such as
the complex electron spectrum model \citep{schl87} or the model with
anisotropic pitch angle distribution of emitting particles \citep{petr01} may
modify the estimates of the IC model quoted above.

Models where the X-rays are generated by nonthermal electron and proton bremsstrahlung
are associated with an unacceptably large energy output of emitting particles
\citep{dog01,petr01}.\\
However, the bremsstrahlung model has not been completely explored because
emission can also be produced by the quasi-thermal electron component which
arises naturally when particles are accelerated from the background thermal
plasma. In this case a part of the spectrum is formed under the influence of
both Coulomb collisions and a run-away flux of accelerated particles
\citep{gur60}. This class of models was developed by \citet{dog00} and
\citet{lia02}, who assumed that the hard X-ray flux from a galaxy cluster is
generated in regions of electron in-situ acceleration from the thermal pool. In
such models the electron distribution function develops an extended transition
populated by quasi-thermal electrons. In this region, which lies between the
thermal and nonthermal parts of the spectrum, the distribution function differs
strongly from the Maxwellian form  because it is not an equilibrium
distribution (and in this sense it is not thermal), but is formed entirely by
the Coulomb collisions (and therefore we cannot define it as nonthermal).

The bremsstrahlung output from the emitting particles is proportional to the
lifetime of their distribution function. For electrons with totally nonthermal
energies, this lifetime is about the timescale on which a single non-thermal
electron suffers ionization losses. In this regime Petrosian's (2001) estimates
of the energy output required to reproduce the hard X-ray excess of Coma is
completely correct. However, for a given energy in the regime where the
spectrum is formed by Coulomb collisions, the lifetimes of the distribution
function and that of a single electron may differ dramatically.
In fact, the lifetime of, e.g., an equilibrium spectrum is much larger than the
characteristic lifetime of a single particle of this spectrum.
Therefore, if the hard X-ray Coma flux is emitted by electrons from the
quasi-thermal transfer region, a specific and more detailed analysis is
necessary for such a situation.

As it was shown by \citet{dog00} and \citet{lia02}, if the plasma temperature is of order
of several keV, then hard X-rays in the energy range $20 - 80$~keV are produced by this
quasi-thermal component of the electron flux. For a fixed radiated luminosity, such as
that of the hard X-ray flux from Coma, the rate of energy input into the electrons can be
lower than that required in nonthermal bremsstrahlung models.  We shall discuss below
this energetic problem in more detail.  We will show, in addition, that this acceleration
process can be tested by looking at the spectral changes in the associated
Sunyaev-Zeldovich (SZ) effect from the intracluster medium.

The problem here described will be analyzed in the present paper under the following
assumptions:
\begin{enumerate}
\item The hard X-ray flux is produced by bremsstrahlung of subrelativistic
electrons;
\item Electrons are in-situ accelerated from background plasma by a stochastic,
Fermi-II type acceleration mechanism;
\item The characteristic time of acceleration is much larger than the
characteristic time of Coulomb collisions of thermal particles, i.e. a small part of
background particles is accelerated. Therefore, the parameters describing the background
plasma change very slowly;
\item This allows to neglect the nonlinear terms in kinetic equations;
\item The particle acceleration is investigated in subrelativistic energy range.
We do not consider, therefore, acceleration of fully relativistic electrons.
\end{enumerate}

The first part of the paper is focused on the energy problem. We will show that
the excess of hard X-ray emission above the thermal spectrum of Coma can be
produced by electron bremsstrahlung and that the necessary energy output of
electrons can be smaller than in the nonthermal bremsstrahlung model. Our
analysis is based on simplified analytical solutions of the kinetic equations.
Nonetheless, beyond the simplified description of the problem, it shows that
the crucial energy problem set by the HXR excess can be definitely alleviated.

In the second part of the paper we derive the characteristics of in-situ acceleration
from the observed spectrum of X-ray emission and we calculate the distortions of the
equilibrium Maxwellian spectrum due to this acceleration. We finally explore whether a
signal of acceleration in the Coma halo can be indicated by the associated
Sunyaev-Zeldovich effect.
In order to compare our results directly with those of Petrosian (2001), we use
here the same cosmological model with $H_0= 60$ km s$^{-1}$ Mpc$^{-1}$.

\section{Kinetic Equation for Electrons}\label{sec:coeffs}

Shock acceleration is usually considered as a candidate for particle production
in the intracluster medium \citep[see, e.g.,][]{kuo03}, and for some clusters,
the available observations suggest that there are shocks \citep[see,
e.g.,][]{marke,fabian} and electrons which might be accelerated by shocks
\citep[see, e.g.][]{bru01, mini01,mini03}. However, extended radio and
(perhaps) X-ray emission cannot be associated with strong shocks since
accelerated electrons are unable to travel large distances from their sources
without loosing much of their energy.

Plasma turbulence is hence considered as a viable model for particle
(re)acceleration in cluster halos. Numerical calculations show that strong
turbulence can be excited in halos \citep[see, e.g.,][]{rick01}. If this
turbulence generates plasma waves, then a slow stochastic process can
accelerate particles through their resonant interactions with the waves. This
process can be described as momentum diffusion. We note that particles can also
be accelerated directly by hydrodynamic turbulent or quasi-periodic flows in a
manner similar to stochastic acceleration by plasma waves \citep[see,
e.g.,][]{webb}.

The evolution of the distribution function of particles which are scattered by
electromagnetic fluctuations is described by the Fokker-Planck equation which
can be transformed to the diffusion type equation by integration over particle
pitch-angles, if scattering is very effective and the distribution function is
quasi-isotropic. For the mechanism of the in-situ acceleration from background
plasma, the equation can be written in the form
\begin{equation}\label{e_k}
 {{\partial f}\over{\partial t}}+{1\over p^2}{\partial\over{\partial
 p}}p^2\left[\left(\frac{dp}{dt}\right)_C f - \left\{D_c(p)+D(p)\right\}{{\partial f}\over{\partial
 p}}\right]=0\,,
\end{equation}
where $(dp/dt)_C$ and $D_c(p)$ describe particle convection and diffusion in the momentum
space due to Coulomb collisions, and $D(p)$ is the diffusion coefficient due to the
stochastic acceleration.
Eq.(\ref{e_k}) can be written, in a dimensionless form, as
\begin{equation}\label{e_kin}
 {{\partial f}\over{\partial \tilde{t}}}-{1\over \tilde{p}^2}{\partial\over{\partial
 \tilde{p}}}\left(A(\tilde{p}){{\partial f}\over{\partial
 \tilde{p}}}+B(\tilde{p})f\right)=0\,,
\end{equation}
where $\tilde{p}=p/\sqrt{mkT}$ is the dimensionless momentum,  $\tilde{t}=t\nu$
is the dimensionless time and  $\tilde{D_p}(\tilde{p})=D_p(p)/(\nu mkT)$ is the
diffusion coefficient. The frequency $\nu$ is
\begin{equation}
 \nu={{2\pi nc^2r_e^2m}\over\sqrt{mkT}} \, ,
\end{equation}
where $r_e=e^2/(mc^2)$ is the classical electron radius.
Here
\begin{equation}
 B(\tilde{p})=\tilde{p}^2\left({{d\tilde{p}}\over{d\tilde{t}}}\right)_i\,,
\end{equation}
and
\begin{equation}
 A(\tilde{p})=B(\tilde{p}){\gamma\over\sqrt{\gamma^2-1}}\sqrt{{kT}\over{mc^2}}
 +\tilde{p}^2\tilde{D}_p(\tilde{p})\,.
\end{equation}
The dimensionless rate of ionization loss has the form
\begin{eqnarray}
 &&\left({{d\tilde{p}}\over{d\tilde{t}}}\right)_i={1\over
 \tilde{p}}\sqrt{\tilde{p}^2+{{mc^2}\over{kT}}}{\gamma\over\sqrt{\gamma^2-1}}\\
 &&\times\left\{\ln\left[{{Emc^2(\gamma^2-1)}\over{h^2\omega_p^2\gamma^2}}\right]
 +0.43\right\}\,, \nonumber
\end{eqnarray}
where $\omega_p$ is plasma frequency and $E(\tilde{p})$ is the particle total energy.
The quasi-steady state solution of Eq.~(\ref{e_kin}), obtained by \citet{gur60}, reads
\begin{equation}
 \label{fgur}
 f=\sqrt{2\over\pi}n(\tilde{t})\exp\left(-\int\limits_0^{\tilde{p}}{{B(v)}\over{A(v)}}dv\right)G(\tilde{p})\,,
\end{equation}
where
\begin{equation}
 G(\tilde{p})={{\int\limits_{\tilde{p}}^\infty[dv/A(v)]\exp\left(\int\limits_0^v[B(t)/A(t)]dt\right)}\over
 {\int\limits_0^\infty[dv/A(v)]\exp\left(\int\limits_0^v[B(t)/A(t)]dt\right)}}\,,
\end{equation}
and $n(t)$ describes slow variations of the background plasma density, consistently with
the assumption of slow acceleration.

Detailed information on the conditions necessary to derive a reliable momentum
diffusion coefficient $D(p)$ are not well determined yet. First of all, its
value  is determined by a spectrum of electromagnetic fluctuations $W(k)$
(where $k$ is the wave-number of fluctuations) which is basically unknown,
though new theoretical treatments of plasma turbulence \citep{verma} or X-ray
observations of clusters \citep[see][]{schuecker} indicate a
Kolmogorov-Oboukhoh type spectrum of turbulence in the large scale range,
between 20 kpc and 2.8 Mpc. Secondly, we do not know in details the ratio
between the energy density of the intracluster plasma, $W_{th}\simeq 1$ eV
cm$^{-3}$, and that of the magnetic field, $U_H$. This ratio is, moreover,
totally unknown in the regions of particle acceleration. For the magnetic field
strength observed in clusters (whose estimates ranges from $\sim 0.1$ to $\sim
10$ $\mu$G, (see e.g. Carilli \& Taylor (2002) for a review), the ratio
$\beta=W_{th}/U_H$ is in the range  $\sim 0.4$ (low $\beta$ plasma) up to $\sim
1000$ (high $\beta$ plasma).
Additional theoretical uncertainties come from the lack of knowledge of the
mechanism through which turbulence is formed in regions of particle
acceleration, whether it is developed by cascade processes (as it is in the new
model of turbulence developed by \citet{gold}), by intermediate turbulence
model \citep[see, also][]{chol} or if it is due to the flux instability when a
flux of particles escaping from acceleration regions excites there MHD
fluctuations due to resonant interaction, as it may occur in the Galactic halo
\citep[see][]{dog94} or near shock fronts \citep{ptuz}. Given all these
uncertainties, we are unable to choose reliable parameters of the kinetic
equations, and we have to resort to a quite general description whose overall
features can be, nonetheless, tested against the available data.

In the simplest case of charge particle scattering, the momentum diffusion
coefficient has the form
\begin{equation}
D_p(p)=D_0p
\label{topt}
\end{equation}
for nonrelativistic particles \citep[see][]{topt}.

For the case of low-$\beta$ plasma the momentum diffusion coefficient  was derived by
\citet{mil92,stein92,mil96,sch98} for resonant particle-wave acceleration in solar corona
and in the interstellar medium of the Galaxy. Its analytical form taken from
\citet{sch98} is
\begin{eqnarray}
 D_p(p)& = & {{\pi(j-1)c_1(s)}\over 4}\left({W_{t}\over U_H}\right) \times \nonumber \\
 & &\Omega(r_Lk_{min})^{j-1}{{v_Ap^2}\over v^2}\ln{v\over v_A}\,,
 \label{c_dif}
\end{eqnarray}
where $r_L$ and $\Omega$ are the Larmor radius and cyclotron frequency of
protons, $k_{min}$ is the minimum wave number of the MHD spectrum, $W_{t}$ is
the total energy of the magnetic fluctuations and $U_H$ is the energy density
of the large scale magnetic field. Here the spectrum of magnetic fluctuations
is supposed to be described by a power-law, $W(k)\propto k^{-j}$, where
$c_1(j)$ is a constant depending on the spectral index $j$.

The momentum diffusion coefficient in a high-$\beta$ plasma is determined by the spectrum
of magnetic turbulence excited by stochastic plasma motion. In the approximation of
strong turbulence, corresponding to high-$\beta$ plasma, its value depends on whether the
accelerated particles are magnetized or unmagnetized in a random magnetic field.
The coefficients of kinetic equations are determined by pair-correlations of
random velocity and random magnetic fluctuations. Note that there is no
resonant interaction in this case. Below we present general equations for the
momentum diffusion coefficient derived by \citet{dog87}. For magnetized
particles the coefficient of momentum diffusion is given by
\begin{equation}
D_p=6\int\limits_{-\infty}^t<VV^\prime><hh^\prime><\nabla\nabla^\prime
hh^\prime>dt^\prime
\end{equation}
where $V$ is the turbulent velocity of the plasma, ${\bf h}={\bf H}/\mid{\bf H}\mid$ is
the random direction of the magnetic field line. The values of $<VV^\prime>$,
$<hh^\prime>$, and $<\nabla\nabla^\prime hh^\prime>$ are the pair-correlations of
turbulent velocity, direction of magnetic field and its derivative, respectively.
For unmagnetized particles one obtains
\begin{equation}
D_p=\frac{2e^2}{3c^2}\int\limits_{-\infty}^\infty <V^2>_{\tau , v\tau}<H^2>_{\tau ,
v\tau}d\tau \, ,
\end{equation}
where
\begin{equation}
<V^2>_{\tau , v\tau}=\int\limits_{-\infty}^\infty \mid
V(k,\omega)\mid^2\exp\left(i(\omega-{\bf kv})\tau\right)d\omega d^3k \, .
\end{equation}
From an estimate of the acceleration time scale, we can derive  the energy density of the
resonant magnetic fluctuations in the case of low-$\beta$ plasma and the correlation
length of turbulence in the case of high-$\beta$ plasma.\\
At the present stage of our knowledge, we cannot prove or disprove the validity
of any form of the diffusion coefficient; however, we notice that \citet{cla01}
found from ROSAT and radio observations that the total magnetic energy content
in clusters is comparable to the total thermal energy content in the same
cluster volume.

To circumvent this problem, we derive here the general characteristics of the
acceleration mechanism  -- i.e., the characteristic time of acceleration (the
dimensionless parameter $\alpha$, see below) necessary to produce the X-ray
excess above the thermal distribution -- from the observed flux of hard X-rays
from Coma, a procedure which is independent of the details of the acceleration
process.

In the following, the spectrum of turbulence is assumed to be a power-law,
$W(k)\propto k^{-j}$, with $j\simeq 2$, i.e. a spectrum that is close to that
derived by \citet{schuecker}. For this fluctuation spectrum the momentum
diffusion coefficient $\tilde{D}(\tilde{p})$ is also a power-law function of
$\tilde{p}$ in a confined range of momenta as follows from
\citet{mil92,stein92}(see.Fig. 5).\\
The energy of accelerated particles
 \begin{equation}\label{wen}
 W_e\propto \int\limits_0^\infty p^4f(p)dp\,,
\end{equation}
diverges for any reasonable power-law form of the diffusion coefficient as
follows from Eq.(\ref{fgur}). In order to avoid this divergence we should make
the natural assumption that there is a cut-off in the fluctuation spectrum for
a wave number ${\bar k}$ that gives the maximum momentum ${\bar p}$ of
accelerated particles. We discuss the expected solution using a simple form of
the momentum diffusion coefficient
\begin{equation}
 {\bar D}(p)= \left\{{\begin{array}{ll}
  \tilde{D}(\tilde{p})&~~~~~~~~~~~~~\tilde{p}<{\bar p}\\
       0&~~~~~~~~~~~~~\tilde{p}>{\bar p}
 \end{array}}
 \right.
 \label{dpp}
\end{equation}
The boundary conditions at $\tilde{p}={\bar p}$ are discussed in the Appendix.

We set the cut-off momentum ${\bar p}$ at an energy ${\bar E}={\bar p}^2/2m
> 80$~keV, above the range of observations, so that it does not affect the
calculated bremsstrahlung spectrum
\footnote{As follows from observations of the X-ray flux from the Galactic disk
\citep[see][]{lebr}, similar processes may accelerate particles up to $\la 100$ keV in
the interstellar medium, although one cannot exclude the possibility that electrons are
accelerated to far higher energies.}.
In our calculations we take the minimum acceptable value of the cut-off energy to be
${\bar E}_k=110$ keV. In the case of resonant  interaction this energy corresponds to the
minimum  wave-number $k_{min}\sim 4.3\cdot 10^{-9}$ cm$^{-1}$ in the spectrum of magnetic
fluctuations $W(k)$.  Therefore, the energy density of resonant waves is
\begin{equation}
W_t=\int\limits_{k_{min}}^\infty W(k)dk \, .
\end{equation}
In the nonrelativistic energy range, eq.(\ref{e_kin}) is simplified significantly and
takes the form
\begin{equation}\label{m_kin}
 {{\partial f}\over{\partial \tilde{t}}}-{1\over \tilde{p}^2}{\partial\over{\partial
 \tilde{p}}}\left[\left({1\over \tilde{p}}+\tilde{D}(\tilde{p}) \tilde{p}^{2}\right){{\partial f}\over{\partial
 \tilde{p}}}+f\right]=0 \, .
\end{equation}
For the qualitative analysis of the problem we consider  the dimensionless diffusion
coefficient in the power-law form $\tilde{D}(\tilde{p})=D(p)/(\nu_1mkT)=\alpha
\tilde{p}^q$ that enable us to get simple analytical solutions of eq.(\ref{m_kin}). Thus,
for $q=1$ the solution of eq.(\ref{m_kin}) has a simple form \citep{gur60}
\begin{equation}
   \label{gur0}
   f(\tilde{p})\propto
   \exp\left({{\pi-2\arctan(\sqrt{\alpha}\tilde{p}^2)}
   \over{4\sqrt{\alpha}}}\right)-1
   \, .
  \end{equation}
Note that here
\begin{equation}
 \nu_1={{4\pi ne^4}\over{(kT)^{3/2}\sqrt{m}}}\ln\Lambda\,,
\end{equation}
where $\ln\Lambda$ is the Coulomb logarithm.

The dimensionless parameter $\alpha$ for Coma can be derived from observational
data. Thus, \citet{dog00} analyzed electron acceleration in the central part of
Coma with the density $n\simeq 10^{-3}$ cm$^{-3}$ while \citet{lia02} derived
parameters of acceleration for the average gas density in the Coma halo,
$n\simeq 10^{-4}$ cm$^{-3}$. It is rather difficult to compare results of these
investigations because different forms of momentum diffusion were used in these
publications.

We notice that the solution (\ref{gur0}) provides an illustrative
oversimplification of the solution of Eqs.(\ref{fgur}) which is useful in order
to get rough quantitative estimates.
We present in the following numerical calculations of Eq.~(\ref{m_kin}) which
show the time variation of the distribution function $f$ under the influence of
stochastic acceleration and Coulomb collisions.

\section{Time Variations of the Spectrum of Accelerated Particles}\label{sec:timevar}

We solve Eq.~(\ref{m_kin}) numerically in order to understand the time evolution of the
distribution function $f$ for a diffusion coefficient
$\tilde{D}(\tilde{p})=\alpha\,\tilde{p}^q$ with $\alpha\approx 0.001$ and $q=1$. As
discussed earlier, $\tilde{D}(\tilde{p})$ must have a cutoff for large enough momentum.
For simplicity, we choose the cutoff momentum to be larger than the maximum momentum we
used in our numerical calculations. In fact, the position of the cutoff momentum does not
significantly affect the evolution of the distribution function at small momenta.

We considered two initial cases:
{\it i)} transformation of the Maxwellian distribution under the
 influence of stochastic acceleration. In this case we can estimate
 the characteristic time for the formation of the particle excess above the
 thermal distribution;
{\it ii)} transformation of the resulting nonequilibrium spectrum under
 the influence of Coulomb collisions only.\\
These calculations allow us to estimate the power of the electron source necessary to
compensate collisional dissipation and to keep the particle excess above the thermal
distribution at the level necessary for the production of the observed hard X-ray
emission in Coma.

In both cases, we calculate the evolution of the electron spectrum up to the
dimensionless time $\tilde{t}=4000$ (which is equal to 1.85~Gyr for Coma, where
$\nu_1\approx 7.2 \times 10^{-14}$ s$^{-1}$). As for the boundary conditions,
we use a free boundary condition at the high-momentum boundary of our
calculations, while at the low momentum boundary we used zero-flux boundary
conditions.

Figure 1 shows the temporal evolution of the distribution function $f$ formed
by the combined effects of stochastic acceleration and thermalization due to
the Coulomb collisions.
The distribution function is normalized to the value at the low momentum boundary. The
dashed curve shown in this figure is the original Maxwellian distribution, and the dotted
curve is the steady state solution (with a cutoff at large momentum). The five solid
curves represent the distribution function at increasing times $\tilde{t}=800, 1600,
2400, 3200$, and $4000$. The figure shows that the distribution function rapidly
approaches the steady state solution for the relatively low momenta of the quasi-thermal
particle regime ($\tilde{p} \la 10$). However, for nonthermal relativistic particles the
stationary state is reached on timescales longer than a cluster lifetime. Therefore, in
the framework of this model it is difficult to expect that subrelativistic electrons
emitting hard X-rays and relativistic electrons emitting radio emission are produced by a
single mechanism of particle acceleration. Therefore, this analysis cannot be extended to
relativistic energies. Our assumption that the maximum energy of the accelerated
electrons is $110$ keV is hence justifiable on the basis of the previous results.
Moreover, as shown by \citet{wolfe} based on a covariant kinetic theory of
electron plasmas, a power-law tail obtainable from direct or stochastic
acceleration of relativistic particles cannot survive for times longer than
$\sim 20$ Myrs because the equilibration time scale for relativistic electrons
is quite short for the case of clusters and hence a thermal distribution is
soon established.
\begin{figure}[h]
\begin{center}
\epsfig{file=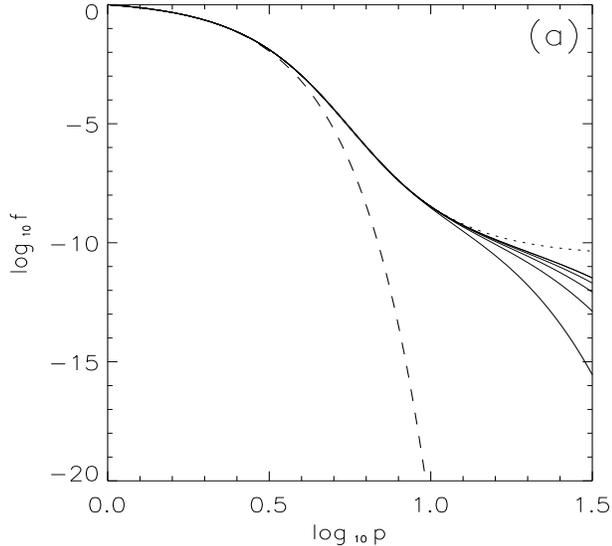,height=8.cm,width=9.cm,angle=0.0}
 \caption{Temporal evolution of the distribution function up to
  $\tilde{t}=4000$ (1.85~Gyr) after acceleration is switched on. The initial
  distribution is Maxwellian.}\end{center}
 \label{gum}
\end{figure}

Now let us assume that the acceleration mechanism is interrupted but the
initial distribution function is described by the nonequilibrium form of
eq.(\ref{gur0}). In this case, the dimensionless equation for the distribution
function $f$ is
\begin{equation}\label{mu_kin}
 {{\partial f}\over{\partial \tilde{t}}}={1\over \tilde{p}^2}{\partial\over{\partial \tilde{p}}}
 \left[{1\over \tilde{p}}{{\partial
 f}\over{\partial \tilde{p}}}+f\right]\,.
\end{equation}
Figure~2 shows the evolution of the electron distribution function, as
described by Eq.~(\ref{mu_kin}), if acceleration is switched off  and the
distribution function is allowed to evolve from its  quasi-steady state form
under the influence of collisional dissipation.
The collisional regime of quasi-thermal particles lies in the range  $5.5 < \tilde{p} <
30$ for $\alpha=0.001$ and $q=1$.\\
We show in Fig.~2 that the dissipation time scale is larger than the single
electron ionization loss time scale near the thermal particle region, because
the distribution function there is of almost equilibrium form, and it increases
in the region of nonthermal momenta, where the lifetime of particles increases
as $\tau_i\propto p^3$.
\begin{figure}[h]
\begin{center}
\epsfig{file=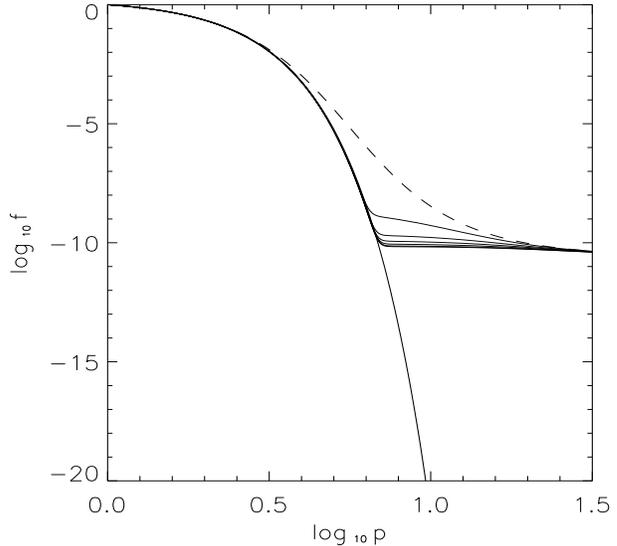,height=8.cm,width=9.cm,angle=0.0}
 \caption{Temporal evolution of the distribution function up to
  $\tilde{t}=4000$ (1.85~Gyr) after acceleration is switched off. The solution (\ref{gur0}) of Eq. (\ref{e_kin}) was taken as an
  initial distribution function.}\end{center}
 \label{ko11}
\end{figure}

\section{The Energetics of Quasi-Thermal Electrons and the Coma hard X-ray Flux}
 \label{sec:energetics}

\begin{figure}[h]
\begin{center}
\epsfig{file=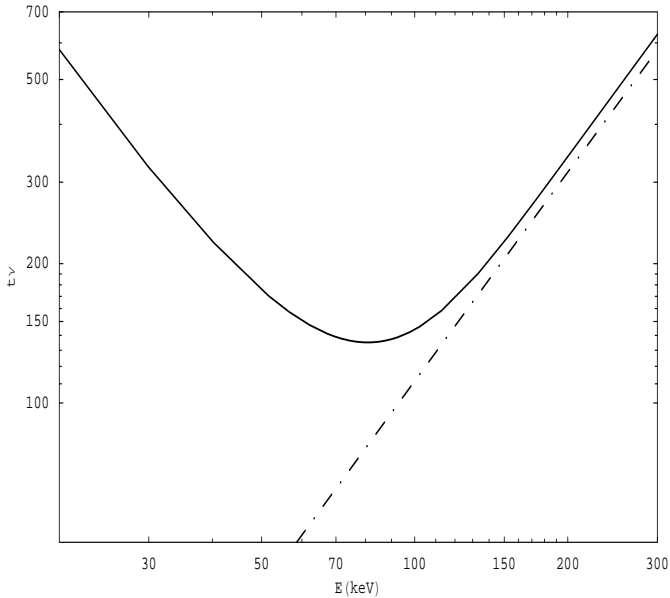,height=8.cm,width=9.cm,angle=0.0}
 \caption{ The lifetime of the electron spectrum (\ref{gur0}) derived from
  Eq.~(\ref{m_kin}) is shown by the solid curve. The lifetime of a single
  electron determined by ionization loss is shown by the dashed-dotted curve.
  Here  the dimension frequency of Coulomb collisions of thermal particles
  $\nu= 7.2\cdot 10^{-14}$ s$^{-1}$ for the Coma parameters $T=8.2$ keV and $n=1.23\cdot 10^{-4}$ cm$^{-3}$.}
  \end{center}
 \label{tnu}
\end{figure}

We now return to the problem of the energetics of the emitting electrons.
 First, we recall Petrosian's criticism of the bremsstrahlung model
\citep[see][]{petr01}. He estimated the yield of bremsstrahlung photons as
$Y\sim (dE/dt)_{br}/(dE/dt)_i\sim 3\times 10^{-6}$ in the energy range 20--80
keV, where $(dE/dt)_i/(dE/dt)_{br}$ is the ratio of ionization to
bremsstrahlung losses. Then for the hard X-ray flux from Coma in this energy
range, $F_x \simeq 4 \times 10^{43}$~erg s$^{-1}$, a large amount of energy,
$F_e\sim F_x / Y \sim 10^{49}$ erg s$^{-1}$ is transferred from the accelerated
electrons to the background plasma by ionization losses. As a result, the
intracluster plasma temperature should rise to a temperature $> 10^8$ K on a
quite short time scale $\sim 3\times 10^7$~yrs.
{We stress here that} these conclusions were obtained under the assumption that
the lifetime of a single electron equals the lifetime of the particle
distribution function. These estimates are correct only in the case that the
electrons are nonthermal, and therefore collisionless. However, the previous
energetic arguments cannot be used in energy ranges where the spectrum is
formed by Coulomb collisions because, as we have shown in
Section~\ref{sec:timevar}, the lifetime of particles differs strongly from {the
lifetime} of the distribution function (see Fig.\ref{tnu}).
For instance, let us consider the lifetime of thermal electrons at an energy of
about 8 keV. Their individual lifetime is about $\sim 10^6$ yr, but the
lifetime of the distribution function at these energies is much longer (almost
infinite) because the distribution function for these energies is almost in
equilibrium. Therefore, the energy supply necessary for the bremsstrahlung
radiation can only be estimated from the corresponding kinetic equations. It
follows that estimates of the energetics  based on the lifetime of single
electrons are not appropriate here, and lead to wrong conclusions.
Figure 3 shows, in fact, the variation of the lifetime of the distribution
function in eq. (\ref{gur0}), as derived from Eq.~(\ref{mu_kin}). From this
figure we see that the lifetimes of the distribution function and of the
particles are equal {to each other} only for high (nonthermal) electron
energies.

In order to estimate the energy supply necessary to support the non-equilibrium
distribution (eq.\ref{gur0}) we use the following kinetic equations in which we include
bremsstrahlung losses
\begin{eqnarray}
 \label{m2_kin}
 &&{{\partial W_e}\over{\partial t}}=\Phi\\
 &&=4\pi VE_e\left({1\over p}{{\partial f}\over {\partial
 p}}+\left(1+p^2\left({{dp}\over{dt}}\right)_{br}\right)f\right)\nonumber\,.
\end{eqnarray}
Here the total number of particles with momentum $\geq p$ in a volume $V$ is
$F_e(p)=V\int^\infty_p f(p)4\pi p^2dp$, and the total electron energy in this volume is
$W_e\simeq E_e F_e$, where the  energy $E_e=kTp^2/2$. Then from Eq.~(\ref{mu_kin}) we
obtain an expression for the rate of change of the energy content of the electrons. {The
number of electrons emitting 50 keV radiation is fixed in order to satisfy the
observations.}

In the case of nonthermal electrons whose spectrum is, e.g., a power-law, the
last term exceeds the others on the right-hand side of Eq.~(\ref{m2_kin}), and
then
\begin{equation}
 \left({{\partial W_e}\over{\partial t}}\right)_{nth}=\Phi_0\simeq 4\pi
 VE_ef\sim {W_e\over\tau_i},
\end{equation}
which is the result discussed by Petrosian (2001), where the electron energy
loss rate is determined by the ionization loss of nonthermal particles. Here
$\tau_i$ is the characteristic time scale of the ionization loss at the energy
$E_e$.

If the particles are thermal, and their spectrum is described by a Maxwellian, then the
first and the third terms on the right-hand side of eq.(20) cancel out so that
\begin{eqnarray}
 \left({{\partial W_e}\over{\partial t}}\right)_{th} &\sim & 4\pi VE_e p^2\left(
 {{dp}\over{dt}}\right)_{br}{{\partial f}\over{\partial p}} \nonumber\\
 &&\sim {W_e\over\tau_{br}}\sim \Phi_0{{(dE/dt)_i}\over{(dE/dt)_{br}}}\,.
\end{eqnarray}
Here, the time-scale $\tau_{br}$ is the characteristic time for bremsstrahlung loss, and
$\tau_{br} \gg \tau_i$. Hence, $(dW_e/dt)_{nth} \gg (dW_e/dt)_{th}$ for the same flux of
bremsstrahlung radiation produced by these electrons.

\begin{figure}[h]
\begin{center}
\epsfig{file=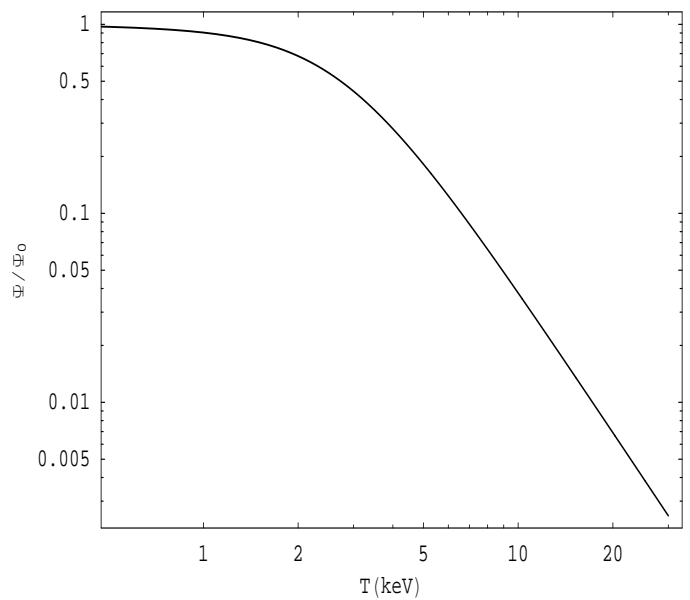,height=8.cm,width=9.cm,angle=0.0}
\caption{The rate at which energy must be supplied to the 50-keV electrons
   generating a fixed bremsstrahlung X-ray flux as a function of
   temperature of background gas, normalized to $\Phi_0$, the rate at
   which energy must be supplied if the emitting electrons are
   non-thermal. For Coma $\Phi_0\simeq 10^{49}$ erg
   s$^{-1}$ \citep{petr01}}.
 \end{center}
  \label{lum}
\end{figure}

We can  estimate the rate of energy supply to the 50-keV electrons that is required to
generate the observed flux of $\sim$50-keV bremsstrahlung X-ray emission from Coma as a
function of different values of the plasma temperature $T$. For simplicity, we consider
the electron spectrum to be of the form given by eq.(\ref{gur0}). When the temperature
$T$ is low, the 50-keV electrons are in the nonthermal particle regime, while for high
plasma temperature these electrons are thermal. From Eqs.~(\ref{gur0}) and (\ref{m2_kin})
we obtain the variation of the associated rate of electron heating as a function of the
background temperature $T$. The required level of heating, normalized to the rate of
heating required for non-thermal particles, $\Phi_0= (dW_e/dt)_{nth}$, is shown as a
function of the temperature in Fig. \ref{lum}, where the parameters defining the
acceleration of particles are those inferred from the X-ray analysis of the Coma cluster.
We find that when the temperature $T$ is low, the 50-keV electrons are nonthermal and --
as expected -- the rate at which they must be heated is independent of the temperature
and is almost equal to $\Phi_0$. However, when the plasma temperature $T$ increases, and
the 50-keV electrons shift into the quasi-thermal regime (where the spectrum is formed by
the Coulomb collisions), the required energy supply to maintain the observed hard X-ray
flux from Coma decreases rapidly as the background plasma temperature increases. For a
temperature $T\sim 8$~keV, the emitting electrons in Coma are quasi-thermal and the
energy supply they require is  one or two orders of magnitude below Petrosian's (2001)
result, even by using a simple qualitative estimate. Accurate quantitative calculations
may give even higher variations. This result may indeed solve the energetic problem
raised by Petrosian (2001): in fact, the quasi-thermal electrons need much less energy in
order to produce the observed HXR bremsstrahlung radiation with respect to the case of
nonthermal particles.

As it is clear from our analysis, our model describes the processe of in-situ
acceleration reasonably well for relatively long time because of the presence of weak
acceleration mechanisms. Therefore, the time variations of both the plasma density and of
its temperature are very slow.
Attempts to investigate a nonlinear phenomenological model of particle acceleration in
Coma were made by \citet{blasi00}. Such an analysis showed that, due to nonlinear
processes, the temperature of plasma increases slowly. However, there are still questions
on whether this model can describe correctly the process of thermalization in the cluster
atmospheres \citep[see discussion by ][]{wolfe}. The role of non-linear effects certainly
needs further analysis which goes beyond the scope of this paper and we will address this
issue elsewhere.

\section{Bremsstrahlung emission of quasi-thermal electrons}
 \label{sec:blung}

We derive here the spectrum of the emitting electrons in Coma from the X-ray data using
the more accurate solution (see eq.\ref{fgur}) of the  kinetic equation eq.(\ref{e_kin}).
We calculate the flux of hard X-ray emission using the equation
\begin{equation}\label{brem}
 F_x={V\over{4\pi d_L^2}}\int\limits_{E_x}^\infty
 nv{{d\sigma_x}\over{dE_x}}N(E)dE\,,
\end{equation}
with cross-section
\begin{eqnarray}
 {{d\sigma_x}\over{dE_x}} &= & {{16}\over 3}{e^2\over{\hbar c^2}}{{r_e^2mc^2}
 \over{EE_x}} \nonumber\\
 & & \times\ln\left({{\sqrt{E}+\sqrt{E+E_x}}\over\sqrt{E_x}}\right)\,,
\end{eqnarray}
\begin{figure}[h]
\begin{center}
\epsfig{file=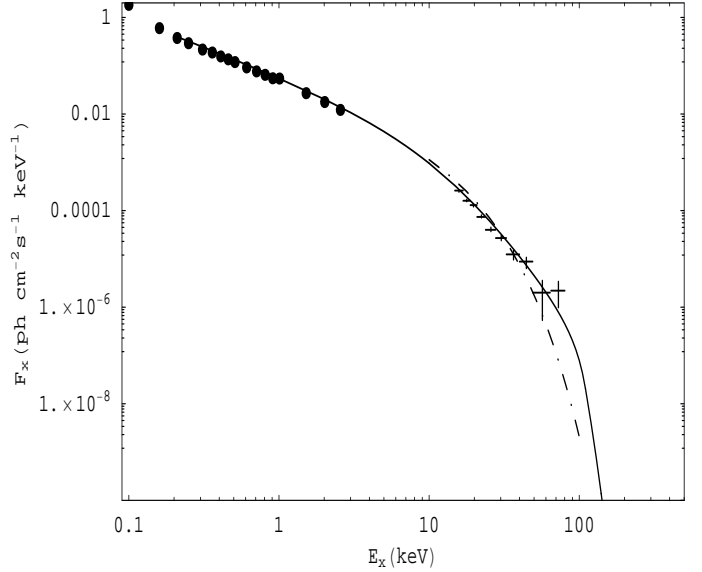,height=8.cm,width=9.cm,angle=0.0}
 \caption{Bremsstrahlung X-ray flux from the Coma cluster produced by
  subrelativistic electrons with a cut-off, $E=110$ keV,
  $T=8.25$ keV. The flux measured with Beppo-SAX
  was taken from \citet{fus99}.}
  \end{center}
 \label{xco}
\end{figure}
The calculated bremsstrahlung spectrum from Coma is shown together with the Beppo-SAX
data in Fig.~\ref{xco}. We consider a the volume of particle acceleration to be $V\simeq
7.74\times 10^{74}$ cm$^{3}$ which is smaller than the whole volume of the Coma halo
inferred from radio and soft X-ray data.
In the general case it is convenient using the simplest form of the diffusion
coefficient in eq.(\ref{topt}). The necessary acceleration time of $\sim $ keV
particles, $\tau_{acc}=p^2/D(p)$, is of the order $\sim 10^{17}$ s,
independently of the acceleration mechanism. This value is slightly lower than
that derived by \citet{dog00} for the central region of Coma where the
intracluster gas has the highest density.

For the case of low-$\beta$ plasma we can go a step further and estimate the
necessary density of resonant wave by using the diffusion coefficients for
electrons and protons taken from \citet{stein92} and \citet{sch98}. We notice
that the momentum diffusion coefficient derived from these equations
corresponds to the general form (\ref{topt}) in the nonrelativistic energy
range.\\
Fig.~6 shows the momentum diffusion coefficient for electrons for the
low-$\beta$ case.
\begin{figure}[h]\begin{center}
\epsfig{file=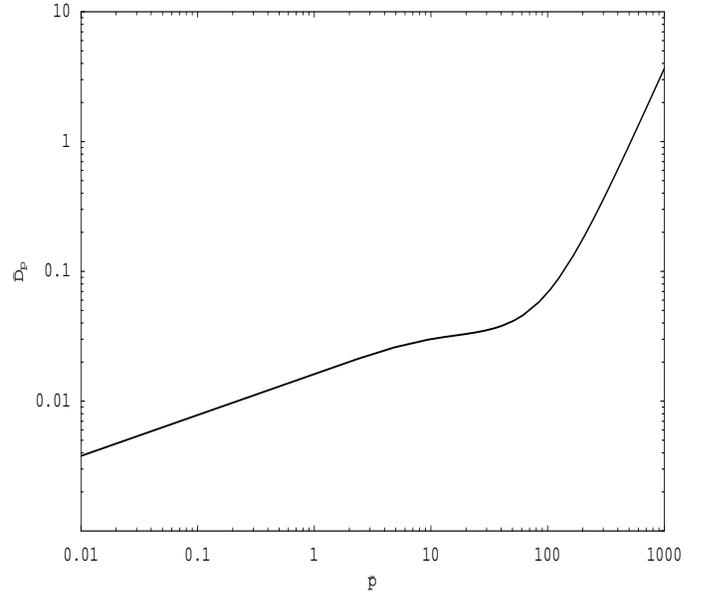,height=8.cm,width=9.cm,angle=0.0}
 \caption{The dimensionless momentum diffusion coefficient $\tilde{D}(p)=D(p)/(\nu mkT)$
 is shown as a function of the dimensionless momentum ${\tilde p}=p/\sqrt{mkT}$.}
  \end{center}
 \label{dpe}
\end{figure}
Only a low energy density in resonant plasma waves, $(W_t/U_H) \simeq 3.8\cdot 10^{-13}$,
is required to fit the Coma HXR data. Here $W_t$ is the energy density of plasma waves
and $U_H=H_0^2/8 \pi$ is the energy density of the large scale magnetic field.

The corresponding electron distribution function is compared with a thermal
spectrum at $T = 8.25$~keV in Fig.~7. The change in the distribution function
is achieved by $\sim 10\%$ of the background electrons being quasi-thermal. It
follows that the energy density of thermal electrons is $W_{thermal}\sim 1.6$
eV cm$^{-3}$ and that of the quasi-thermal electrons is $W_{nonthermal} \sim
0.56$ eV cm$^{-3}$, thus yielding the ratio $W_{nonthermal}/W_{thermal} \approx
0.35$ for Coma.
\begin{figure}[h]
\begin{center}
\epsfig{file=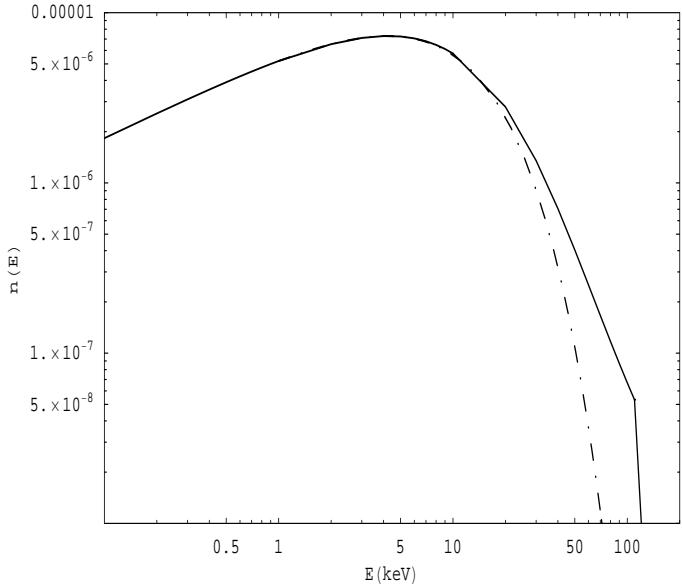,height=8.cm,width=9.cm,angle=0.0}
 \caption{Spectrum of subrelativistic electrons with cut-off, ${\bar p}=110$
  keV, and $T=8.25$ keV derived from the X-ray flux from the Coma
  cluster (solid curve). The thermal spectrum is shown by the
  dashed-dotted curve.}
  \end{center}
 \label{en_e}
\end{figure}

\section{The spectrum of accelerated protons}
 \label{sec:pblung}

Protons are generally disregarded in the calculation of bremsstrahlung emission because,
having low rates of energy loss, they leave the emitting region before they lose a
significant fraction of their energy. On the other hand, if protons escape from a
radiating region relatively slowly, they can produce a bremsstrahlung flux whose value is
comparable with that of the electrons. An analysis of the bremsstrahlung radiation
emitted by subrelativistic protons (inverse bremsstrahlung) in the Coma cluster
\citep{dog01} showed that protons are able, in fact, to generate the observed hard X-ray
emission. These X-rays are accompanied by the excitation of background nuclei which could
be detectable through the development of prominent carbon and oxygen gamma-ray lines.
\citet{dog01} estimated the expected flux of these lines. Recently, \citet{iyud} have
found tracers of this gamma-ray line emission towards the Coma and Virgo clusters at the
expected level. The energy deposition by subrelativistic protons estimated by
\citet{iyud} is of the order of $\sim 8 \times 10^{48}$ erg s$^{-1}$, which matches the
required rate if protons produce the observed hard X-ray flux \citep{dog01}. However,
this process involves subrelativistic protons and so it faces the same problem of
energetics and plasma heating as we discussed in Section~\ref{sec:energetics}.

The theoretical approach we have taken in the present paper allows us to estimate the
density of electrons and protons accelerated from the   background pool using the kinetic
parameters derived from the spectrum   of plasma waves, and hence to understand whether
protons are important.
The momentum diffusion coefficient for protons was taken  in the form of
Eq.(\ref{c_dif}).
\begin{figure}[h]
\begin{center}
\epsfig{file=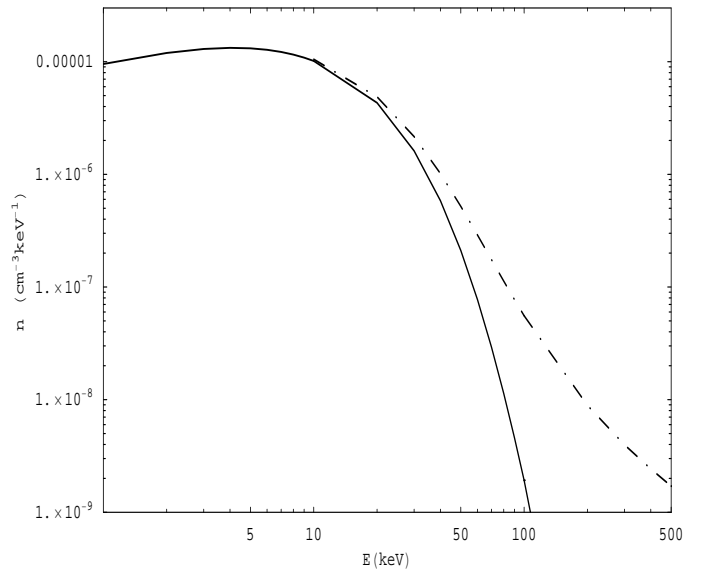,height=8.cm,width=9.cm,angle=0.0}
 \caption{The spectrum of subrelativistic electrons (dashed-dotted curve)
  and protons (solid curve) in the Coma cluster
  halo are shown as a function of the particle energy.}
  \end{center}
 \label{p-e}
\end{figure}
In Fig.~8 the spectrum of protons is shown for the parameters derived in
Section~\ref{sec:blung}. It can be seen that stochastic acceleration is
ineffective for subrelativistic protons  (note that a similar conclusion for
the Galaxy was also obtained by \citet{sch98}). In order to produce a flux of
proton bremsstrahlung comparable with that of electrons, for example at a
photon energy 10~keV, the density of 20-MeV protons should be about the same as
the density of 10-keV electrons. As shown Fig.~8, this condition is not
fulfilled, and so we conclude that the hard-X-ray emission from the Coma
cluster cannot be ascribed to proton bremsstrahlung.

\section{The Sunyaev-Zeldovich effect produced by in-situ accelerated particles}
 \label{sec:sz}

In the context of the present study it is important to find observational
resources which can provide independent evidence for the subrelativistic
electron population that we consider to be responsible for the hard X-ray
emission from clusters of galaxies. One possibility is to use a detailed
analysis of the  inverse Compton scattering of CMB photons off the population
of sub-relativistic electrons, the SZ effect described by Zeldovich and Sunyaev
\citep[see][]{zeld}. The amplitude and the spectrum of this effect depend on
the distribution function of the electrons in the intracluster medium. As a
result of the inverse Compton scattering, the spectrum of the CMB radiation is
shifted to higher frequencies when observed along the line of sight through the
intracluster medium \citep[see for a general review][]{birk99}. The spectral
distortion of the microwave background can be calculated for any electron
distribution function, and not only for the standard Maxwellian spectrum
adopted in most discussions of the Sunyaev-Zeldovich effect: hence, we can
calculate the Sunyaev-Zeldovich effect as due to a nonthermal component of the
electron distribution in the supra-thermal{ \citep[see, e.g.,][]{bla00,enss00}
and relativistic \citep[see][]{shi02,cola03} energy ranges.}
To this aim, we use here the electron distribution function derived from the
X-ray data to calculate the Sunyaev-Zeldovich effect using the formalism
presented in \citet{reph} and \citet{birk99}. The change in the radiation
temperature $\Delta T(\nu)$ at frequency $\nu$ is given by
\begin{equation}
 {\Delta T(\nu) \over T_{rad}} = {{(e^x-1)^2}\over{x^4e^x}} {{\Delta
  I(\nu)}\over I_0}\,,
\end{equation}
where $x=h\nu/kT_{rad}$ is the dimensionless frequency, $\Delta I$ is the
scattering-induced change in the specific intensity of the cosmic microwave background
spectrum at frequency $\nu$, $T_{rad}=2.73$ K is the temperature of the microwave
background radiation, and the scale of the specific intensity is
\begin{equation}
 I_0={{2(k T_{rad})^3}\over{(hc)^2}} \, .
\end{equation}
The CMB temperature variation  due to the SZ effect is
\begin{eqnarray}
 &&\Delta T(\nu)=T_{rad}\tau{{(e^x-1)^2}\over{e^xx}}\int\limits_{-\infty}^\infty dsP_1(s)\\
 &&\times
 \left({{\exp[-3s)]}\over{\exp[xe^{(-s)}]-1}}-{1\over{\exp[x]-1}}\right)\, , \nonumber
\end{eqnarray}
where $\tau=\sigma_{Th} \int dl n_e$ is the optical depth along a line of sight of length
$l$ through electrons of density $n_e$, with $\sigma_{Th}$ being is the Thomson
cross-section,
and
\begin{equation}
 P_1(s)=\int_{\beta_{min}}^1p_e(\beta)P(s,\beta)\,d\beta\,
\end{equation}
is the photon redistribution function calculated in the limit of single scattering
(appropriate here for low values of $\tau$), with
\begin{equation}
 \beta_{min}={{e^{|s|}-1}\over{e^{|s|}+1}}\,.
\end{equation}
The quantity $p_e(\beta)$ is the normalized electron spectrum  given as a
function of the normalized velocity $\beta =v/c$,
\begin{equation}
 \int\limits_0^1p_e(\beta )d\beta=1\, .
\end{equation}
The function $P(s,\beta)$ describes the logarithmic frequency ratio caused by a single
electron/photon scattering, and it is given in \citet{birk99}.  The general description
of the non-thermal SZ effect for multiple scattering, general values of $\tau$ and
multiple electron distribution can be found in \citet{cola03}.
For practical purposes,  we transform the electron distribution function from its
momentum representation (Fig.~7) to its $\beta$ representation (Fig.~9). The suprathermal
excess of electrons which is evident in these figures is compensated by a reduction in
the number of electrons at lower energies, but this reduction is a small fractional
change of the thermal electron number, and so it is neglected in Figs.~7 and ~9.
\begin{figure}[h]
\begin{center}
\epsfig{file=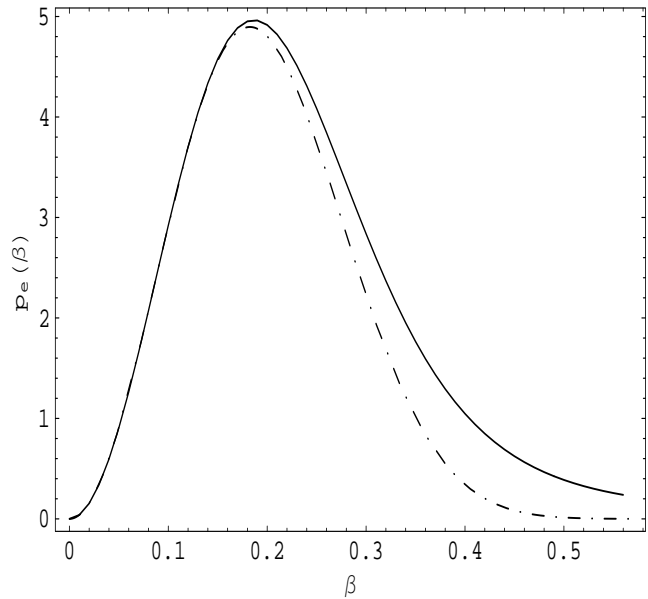,height=8.cm,width=9.cm,angle=0.0}
 \caption{The dimensionless distribution function for subrelativistic
  electrons with cut-off ${\bar p}=110$ keV and $T=8.25$ keV derived
  from the X-ray flux from the Coma cluster (solid curve). The pure
  thermal spectrum is shown by the dashed-dotted curve.}
  \end{center}
 \label{ebe}
\end{figure}

\begin{figure}[h]
\begin{center}
\epsfig{file=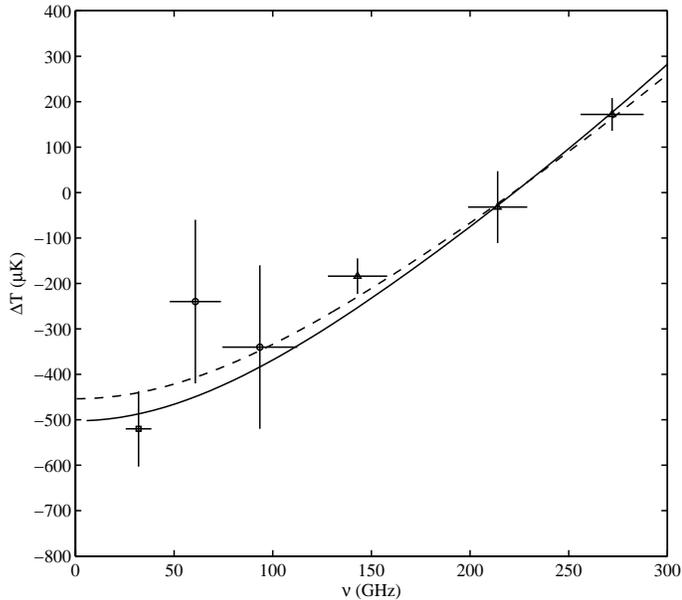,height=8.cm,width=9.cm,angle=0.0}
 \caption{The Sunyaev-Zeldovich effect produced by thermal electrons
  in the Coma cluster region, with temperature $T=8.25$ keV and
  optical depth $\tau=5.3\times 10^{-3}$ (dashed curve). The solid curve
  shows the Sunyaev-Zeldovich effect due to thermal electrons from the
  Coma halo plus thermal and quasi-thermal electrons from regions of
  particle acceleration which occupy a small fraction of the Coma halo
  and have optical depth $9\times 10^{-4}$).}
  \end{center}
 \label{sz}
\end{figure}

\begin{figure}[h]
\begin{center}
\epsfig{file=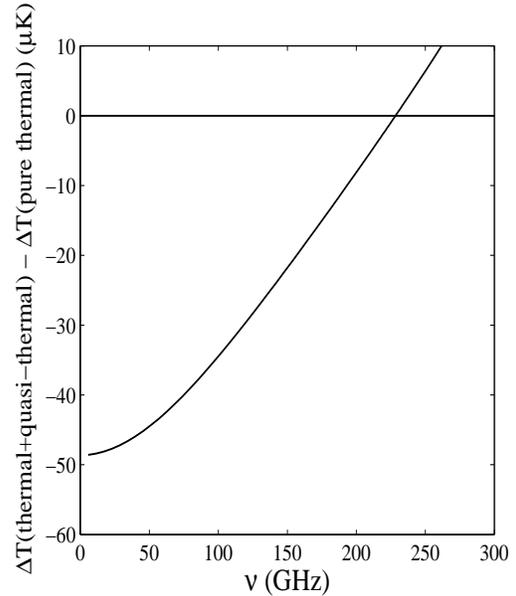,height=8.cm,width=9.cm,angle=0.0}
 \caption{Difference between the pure thermal and the total (thermal
  plus quasi-thermal) Sunyaev-Zeldovich effect.}
  \end{center}
 \label{dsz}
\end{figure}

We can now calculate the Sunyaev-Zeldovich effect from the regions where the acceleration
takes place. Here we should take into account that, as follows from the soft X-ray data,
the volume of the acceleration region, $V_{acc} = 7.7 \times 10^{74} \ \rm cm^3$,  is
comparable with the Beppo-SAX estimate of the emitting volume $V_{em} \leq 1.7 \times
10^{75} \ \rm cm^3$. The total optical depth of the thermal gas in the Coma cluster is
not precisely known. \citet{batt} estimated the total optical depth of Coma as
$\tau_{th}=(5.35\pm 0.67)\times 10^{-3}$ and \citet{dep} estimated the depth as
$\tau_{th}=(4.1\pm 0.9)\times 10^{-3}$. Moreover, some contribution to the optical depth
may be due to an extended halo of cooler gas, with a temperature $0.6 - 1.3$ keV and
density of $10^{-4} - 10^{-3}$ cm$^{-3}$ \citep[see][]{nev}.

We calculate here the Sunyaev-Zeldovich effects from the thermal volume of the Coma halo
in the absence of particle acceleration (the dashed curve in Fig.~10) and in its presence
(the solid curve in Fig.~10). In these calculations we assumed that the temperature and
the optical depth of the thermal electron population are $T=8.25$~keV and $\tau_{th} =
5.3\times 10^{-3}$, as derived from X-ray observations. Such an estimate assumes that the
hot plasma occupies the whole volume of the Coma halo. Assuming that the volume where
particles are accelerated is $V_{acc} =7.7\times 10^{74}$ cm$^3$, we estimate an optical
depth of the acceleration region (i.e., a part of the total optical depth where the
electron spectrum is distorted by the acceleration) which is $\tau_{acc} \approx 9\times
10^{-4}$.

The difference between the pure thermal and the total (thermal plus quasi-thermal)
Sunyaev-Zeldovich effects is shown in Fig.~11. This figure shows that  the presence of
sub-relativistic electrons in Coma produces a further temperature decrement at $\nu \la
230$ GHz and a further temperature increase at higher frequencies. At the same time, the
additional pressure (or energy density) contributed by the sub-relativistic electrons
produce an increase in the frequency of the zero point (i.e., the frequency at which
$\Delta T = 0$) of the SZ effect, a value which is uniquely determined by the overall
pressure $P_{tot}= P_{th} + P_{subrel}$, of the electron population (see Colafrancesco et
al. 2003). Also the overall amplitude of the temperature decrement due to the
subrelativistic electrons is proportional to their pressure since $\Delta T \propto \int
dl P_{subrel}$.

The change in the Sunyaev-Zel'dovich effect due to the accelerated electrons produces a
temperature decrement  $\Delta T \sim 50-40 \ \rm \mu K$ in the frequency range  $50 -
100$~GHz.
However, the uncertainty of the available SZ data for Coma does not allow yet to set
definite constraints on the model presented in this paper. Nonetheless, with the new
generation of telescopes for observing the Sunyaev-Zel'dovich effect, such a signal
should be detectable. The  major difficulty in seeing this departure from the thermal
Sunyaev-Zel'dovich effect is the presence of background fluctuations in the microwave
background radiation. The expected background anisotropy on degree angular scales of
interest for the Coma cluster is $\sim 40 \ \rm \mu K$ in this frequency range, and has a
flat spectrum which is not readily distinguished from the gently-curving spectrum seen in
Figs.~10-11.
However, the use of multi frequency observations increases the possibility of detection
of the SZ effect associated to sub-relativistic particles since the relative amplitudes
of the thermal and subrelativistic SZ effects change with the frequency over the whole
range ($\sim 30 - 300$ GHz) accessible to SZ experiments.
Some possibility of detecting the distorted spectrum of the Sunyaev-Zel'dovich effect in
the presence of background fluctuations also exists through a detailed spectral
measurement of the cluster over a smaller region to reduce the background fluctuation
signal (although this won't reduce the amplitude of the kinematic Sunyaev-Zel'dovich
effect, which has the same spectrum as the background fluctuations).
Measurements of this type are challenging, but should become possible as the
sensitivity of bolometer arrays in the mm and sub-mm bands increases, provided
that accurate cross-calibration in the different bands is achieved
\citep{birklanc}.

The possibility to measure the SZ effect from the population of sub-relativistic or
relativistic particles could provide a way to estimate the overall pressure of such
particle population and, in turn, to constrain the energy spectrum taken up by such
particles in the cluster atmosphere. Even if this demands a large experimental effort,
the result will undoubtedly shed additional, and maybe crucial, light on the nature of
the acceleration mechanism.

\section{Conclusions}
 \label{sec:conc}

We have analyzed in this paper the process of particle acceleration from a
background plasma acting through magnetic fluctuations generated by
intracluster turbulence with a specific application to the problem of the
origin of hard X-ray emission from the Coma cluster. From the equations
describing the influence of Coulomb collisions, we derived the entire electron
distribution function from the thermal to the high-energy non-thermal regime.
Our analysis allowed us to estimate the energy supply necessary for
bremsstrahlung to be responsible for the hard X-ray emission. For nonthermal
electrons we confirm that the bremsstrahlung efficiency is low, which makes it
almost impossible to regard this electronic component as the source of the hard
X-ray excess in Coma. This result is in complete agreement with earlier
conclusion (see Petrosian 2001) on the inapplicability of the nonthermal
bremsstrahlung interpretation. However, we have also shown here that the
bremsstrahlung efficiency increases significantly if the emitting electrons
belong to the extended transfer regime between the thermal and nonthermal parts
of the electron distribution function. This quasi-thermal regime is formed
naturally when emitting particles are accelerated from the background plasma.
In the specific case of Coma, we found that the total energy loss rate of the
quasi-thermal electrons that emit the HXR radiation in the $20 - 80$~keV range
is almost two orders of magnitude lower than for nonthermal particles.
This result may solve both the problem of the origin of the HXR emission in
Coma and of the excessive heating of the cluster gas in the bremsstrahlung
interpretation of the HXR excess.

We have further shown that the distribution function of quasi-thermal electrons that we
derived implies significant distortion of the thermal Sunyaev-Zeldovich effect from the
Coma cluster. Although this additional signal is at the level of $\sim 10\%$ of the
amplitude of the thermal Sunyaev-Zel'dovich effect and therefore its observation will be
challenging from the experimental side, its definite detection will nonetheless be able
to provide a stringent test of our theoretical model.

\acknowledgements{The authors thank the Referee whose comments helped us to improve the
paper. VAD is grateful to Anisia Tang and Boris Klumov for kind help to perform numerical
calculations. CMK is supported in part by the National Science Council of Taiwan grants
NSC-92-2112-M-008-046 and NSC-93-2112-M-008-017. PHK is supported in part by the National
Science Council of Taiwan grant NSC-93-2112-M-008-006. CYH is supported in part by the
National Science Council of Taiwan grant NSC-93-2112-M-008-016. WHI is supported in part
by the National Science Council of Taiwan grants NSC-93-2112-M-008-006 and
NSC-93-2752-M-008-001-PAE. VAD and DAP are partly supported by the grant of a President
of the Russian Federation "Scientific School of Academician V.L.Ginzburg".}

\appendix

\section{kinetic equation}

In the nonrelativistic energy range when $E\gg kT$ ($p \gg 1$), the kinetic equation for
$p<{\bar p}$ and $p>{\bar p}$ can be significantly simplified and written in the form
respectively
\begin{equation}\label{mn_kin}
 {{\partial f}\over{\partial \tilde{t}}}-{1\over \tilde{p}^2}{\partial\over{\partial
 \tilde{p}}}\left[\left({1\over \tilde{p}}+\alpha \tilde{p}^{q+2}\right){{\partial f}\over{\partial \tilde{p}}}+f\right]=0\,,
\end{equation}
\begin{equation}\label{mmkin}
 {{\partial f}\over{\partial \tilde{t}}}-{1\over \tilde{p}^2}{\partial\over{\partial
 \tilde{p}}}\left({1\over \tilde{p}}{{\partial f}\over{\partial \tilde{p}}}+f\right)=0\,,
\end{equation}
with the boundary conditions corresponding to continuity of the function
\begin{equation}
 f({\bar p}, t)={\bar f}({\bar p}, t) \label{bond1}\,,
\end{equation}
where ${\bar f}$ is the solution of Eq.~(\ref{mn_kin}) for the unknown flux of
particles, $-S_0$, running away into the acceleration region
\begin{eqnarray}
 \label{gursol}
 &&{\bar f}(\tilde{p},\tilde{t})=\sqrt{2\over
 \pi}n(\tilde{t})\exp\left[-\int\limits_0^{\tilde{p}}
 {{B(v)}\over{A(v)}}dv\right]\\
 &&-S_0\int\limits_0^{\tilde{p}}{{dv}\over{A(v)}}\exp\left[-\int\limits_v^{\tilde{p}}{{B(t)}\over{A(t)}}dt
 \right]\,. \nonumber
\end{eqnarray}

The flux of particles running into the acceleration region derived from
Eq.~(\ref{gursol}) is
\begin{equation}
 \label{sp}
 S(\tilde{p})=-S_0\sqrt{2\over\pi}\int\limits_0^{\tilde{p}} v^2\exp\left(-v^2/2\right)dv
\end{equation}
and this changes from $S=0$ for $\tilde{p}=0$ to $S=-S_0$ for $p\approx
\tilde{p}_M=\alpha^{-1/(q+3)}$.  Above the momentum $\tilde{p}_M$, the function
$f$ cannot be described by a Maxwellian distribution because the equilibrium
conditions are violated by the run-away flux $S$, but the spectrum is still
formed by Coulomb collisions. For energies above
$\tilde{p}_{inj}=\alpha^{-1/(q+1)}$, the spectrum is nonthermal because Coulomb
collisions are unimportant particle interactions while plasma waves dominate.

It was shown by \citet{dog00} and \citet{lia02} that two excesses above the thermal
Maxwellian spectrum are formed in the range $\tilde{p}>\tilde{p}_M$. When
$\tilde{p}_M<\tilde{p}<\tilde{p}_{inj}$, the excess is formed by Coulomb collisions (the
{\it collisional} regime of quasi-thermal particles), and one can imagine the spectrum
there as a distorted Maxwellian function. For $p>p_{inj}$ the spectrum is formed by
particle interactions with plasma waves (the {\it collisionless} regime of nonthermal
particles). For these energies the particle spectrum can be described as power-law over
an extended energy range.
We also impose a natural boundary condition for $\tilde{p}=\infty$,
\begin{equation}
 f(\infty)=0\,.
\end{equation}

In order to derive the unknown constant $S_0$ we should match solutions of
Eqs.~(\ref{e_kin}) and (\ref{mmkin}) for $\tilde{p}={\bar p}$.

It is convenient to reduce Eq.~(\ref{mmkin}) to the form
\begin{equation}
 \label{eqz}
 {{\partial^2 z}\over{\partial \zeta^2}}-{{\partial z}\over{\partial
 t}}=-z{1\over{4\sqrt{\xi}}}\left({9\over 16}{1\over \xi^2}-1\right)\,
\end{equation}
by introducing the variable $\zeta$,
\begin{equation}
 \zeta= 2^{1/4}  {4\over 5}\left({\tilde{p}^2\over 2}\right)^{5/4}\,,
\end{equation}
and the function $z$,
\begin{equation}
 f=z\left({{\eta^2}\over\sqrt{\ln(1/\eta)}}\right)^{1/4}\,,
\end{equation}
where
\begin{equation}
 \eta = \exp(-\tilde{p}^2/2)\,.
\end{equation}
and
\begin{equation}
\xi =\left({{5\zeta}\over{2^{9/4}}}\right)^{4/5}
\end{equation}
For large $\zeta$ (or equivalent $p$ or $\xi$) values we can neglect the right-hand side
in Eq.~(\ref{eqz}), and the equation takes the well-known one-dimensional diffusion
equation form
\begin{equation}\label{d1}
 {{\partial^2 z}\over{\partial \zeta^2}}- {{\partial z}\over{\partial t}}
 \approx 0\,,
\end{equation}
with boundary condition (\ref{bond1}).
As an initial condition we put
\begin{equation}
 z(\zeta)=0 ~~~~~~~~~\mbox{at}~~~~~t=0,
\end{equation}
i.e., we assume that there were no particles in the momentum range
$\tilde{p}>{\bar p}$ at $\tilde{t}=0$.

The solution of Eq.~(\ref{mmkin}) can be presented with the well-known Green
function $G(x,x_0~|~t,\tau)$ for the diffusion equation
\begin{eqnarray}
 &&G(x,x_0~|~\tilde{t},\tau)=\\
 &&
 {1\over{2\sqrt{\pi}}} {1\over{(\tilde{t}-\tau)^{1/2}}}
 \exp\left[-{{(x-x_0)^2}\over{4(\tilde{t}-\tau )}}\right]\,.
 \nonumber
\end{eqnarray}
Then the function $f$ for Eq.~(\ref{d1}) with the boundary conditions
(\ref{bond1}) has the form
\begin{eqnarray}\label{maxst}
 &&f(\tilde{p},\tilde{t})={\bar f}({\bar p})\left({{{\bar p}~\exp({\bar p}^2)}
 \over{\tilde{p}~\exp(\tilde{p}^2)}}\right)^{1/4}\label{sol}\\
 &&\times\left[1-\Phi \left({{(\zeta-{\bar\zeta})}\over{2\sqrt{t}}}\right)
 \right]\,, \nonumber
\end{eqnarray}
where $\Phi (x)$ is the error function. We see that for $t\rightarrow\infty$
the solution (\ref{sol}) tends to the stationary distribution. In this limit
$\partial f/\partial \tilde{t}~\ll 1$ and our neglect of the right-hand side of
Eq.~(\ref{eqz}) is not valid. Therefore in the stationary distribution limit,
the distribution function $f$ tends to the equilibrium Maxwellian function for
$\tilde{p}>{\bar p}$ when $\tilde{t}\rightarrow \infty$,
\begin{equation}\label{fmco}
 f(\tilde{p},\tilde{t}=\infty)=C(\tilde{t})\exp\left[-{\tilde{p}^2\over 2}\right]\,.
\end{equation}

The constant $C$ and the unknown flux $S_0$ can be defined from the boundary
condition (\ref{bond1}) and the normalization condition
\begin{equation}
 \int\limits_{\bar p}^\infty f(\tilde{p})\tilde{p}^2 d\tilde{p}\simeq S_0t\,.
\end{equation}
However, as we see from Eq.~(\ref{maxst}), in the stationary case we have an
exponential cut-off of the distribution function $f$ for $\tilde{p}={\bar p}$,
which gives an approximate equation for $S_0$ of the form
\begin{eqnarray}\label{s0}
 &&S_0=\sqrt{2\over\pi}n(\tilde{t})\\
 &&\times\left(\int^{\bar p}_0{{dv}\over{A(v)}}
   \exp\left[-\int\limits_0^v{{B(t)}
 \over{A(t)}}dt\right] \right)^{-1}\,, \nonumber
\end{eqnarray}
where $n(t)$ is the density of background plasma, which decreases slowly with
time.


\end{document}